\documentclass[fleqn,10pt,a4paper]{wlscirep}
\usepackage{graphicx,amsmath,amssymb,wasysym,verbatim,color,pdfpages}

\title{Competition and dual users in complex contagion processes}
\author[1,2$*$]{Byungjoon Min}
\author[2,$\dagger$]{Maxi San Miguel}
\affil[1]{Department of Physics, Chungbuk National University, Cheongju, Chungbuk
28644, Korea}
\affil[2]{IFISC, Instituto de F\'isica Interdisciplinar y Sistemas Complejos
(CSIC-UIB), Campus Universitat Illes Balears, E-07122 Palma de Mallorca, Spain}
\affil[*]{bmin@chungbuk.ac.kr}
\affil[$\dagger$]{maxi@ifisc.uib-csic.es}
\date{\today}

\begin{abstract}
We study the competition of two spreading entities, for example innovations, in complex
contagion processes in complex networks.
We develop an analytical framework and examine the role of dual users, i.e. agents using
both technologies. Searching for the spreading transition of the new innovation and the
extinction transition of a preexisting one, we identify different phases depending on network
mean degree, prevalence of preexisting technology, and thresholds of the contagion process.
Competition with the preexisting technology effectively suppresses the spread of the new
innovation, but it also allows for phases of coexistence. The existence of dual users
largely modifies the transient dynamics creating new phases that promote the spread of
a new innovation and extinction of a preexisting one. It enables the global spread
of the new innovation even if the old one has the first-mover advantage.
\end{abstract}

\begin{document}
\flushbottom
\maketitle
\thispagestyle{empty}
\maketitle

\section*{Introduction}

The spread of rumors, fads, innovations or new technologies in a global scale occurs more
frequently than before due to the intensive use of new information and communications
technologies \cite{geroski,karsai}. In order to
model and understand the mechanisms of spreading phenomena~
\cite{schelling,granovetter,watts,centola,gai,dodds,christakis,singh,ruan},
quantitative approaches to contagion processes have become important.
A pioneer work about diffusion of collective behavior \cite{granovetter} and
its variant adapted for the spread on complex networks \cite{watts} proposed
a model of ``complex contagion'' \cite{macy,weng-vir,monsted,min-maxi} incorporating
a threshold mechanism. This is often called the threshold model in which
each agent requires a sufficiently large fraction of neighbors that have already adopted
a spreading new technology to adopt it. Therefore, neighbors affect collectively (group
interaction) the probability of adoption rather than independently as in the ``simple
contagion'' of well known epidemic models \cite{bailey}. Research on spreading by complex
contagion sheds light on fundamental aspects of collective phenomena, nonlinearity in
diffusion, cascading dynamics, and first order phase transitions in complex networks.
It has focused on aspects such as seed effect \cite{gleeson-seed}, clustering \cite{melnik-tree},
modularity \cite{centola,galstyan,gleeson-modular,nematzadeh}, multiplexity \cite{brummitt,lee},
multi-stage contagion \cite{melnik}, and interaction with simple contagion \cite{aga} or
coordination processes~\cite{lugo}.

Models of complex contagion typically assume that there exists no preexisting technology
that competes with a new one, so that the new technology spreads in a system of susceptible
agents~\cite{karsai,watts,centola}. However, it is common that a preexisting technology
is used by some fraction of the agents in the system.
Therefore, the spread of a new competing technology involves the decline, coexistence
or extinction of the actual predominant technology \cite{messenger,sns,cyworld}.
Competition between ideas or products \cite{kocsis,newman,ahn,karrer} is an essential
ingredient in the modeling of contagion process. While there exist several studies of
simple contagion with cooperative epidemics \cite{chen,cai,choi}, competing
epidemics \cite{sahneh,weng}, and interacting epidemics on multi-layer
networks \cite{granell,min,sanz}, complex contagion with interactions among multiple
spreading entities has not been, so far, addressed in detail.

A most clear example of competition in technology spreading is that of new social
networking services in top of a preexisting technology~\cite{messenger,sns}.
A new service competes with preexisting ones in the market share. The initial
technology can take first-mover advantage such as preemption,
technological leadership, and switching cost to other ones, and thus the
spread of the new technology is effectively hindered. Therefore, one
might expect that when an existing product occupies a dominant position
in the market, the global spread of a newly launched product of similar characteristics
is hardly successful.
However, it is not rarely observed that a late technology
successfully spreads into population, such as the success of new online social
networking services \cite{sns}. These observations call for an understanding of the
competition dynamics in complex contagion processes. In this paper we address this
question considering the spreading of a new technology competing with a preexistent one.
An important fact in this context is that some agents use multiple technologies at
the same time. For instance, it is reported that 52 \% of people use multiple different
social networking services because different neighbors use different software \cite{sns}.
We thus explore the role of the agents using  at the same time the old and new technologies.
The consideration of these ``dual users'' is also motivated by analogy with bilingual agents
on problems of language competition \cite{castello,carro,amato} where they modify in
essential ways the competition dynamics. Within a rich phenomenology of spreading and
extinction transitions, as well as phases of coexistence, we find that the existence
of dual users facilitates the spread of a new technology and the decline of the
preexisting one. Dual users act as catalysts of extinction transitions and coexistence phases.
Therefore, the conflict between the first-mover advantage and success of a late mover
can be resolved by introducing ``dual users'' which promote the spread of
a late mover.

\section*{Model}
\begin{figure}[t]
\includegraphics[width=\linewidth]{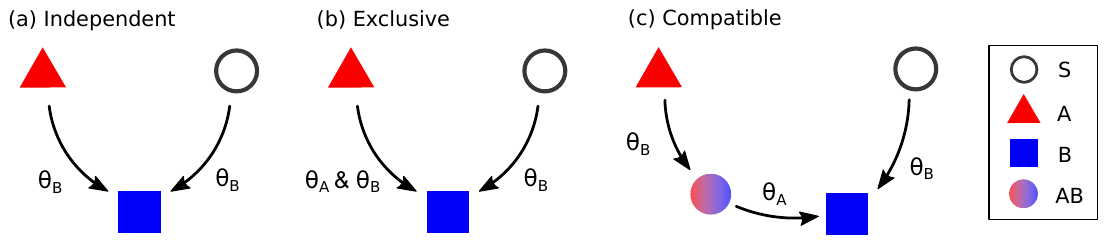}
\caption{
Schematic illustration of (a) independent, (b) exclusive, and (c) compatible
models. Open circle, triangle, square, and filled circle respectively represent
susceptible $S$, $A$, $B$, and dual user $AB$.
}
\label{fig:model}
\end{figure}

Here we consider a model of complex contagion of a new product $B$ with threshold
mechanism~\cite{granovetter,watts} on a network with $N$ nodes in which a preexisting
technology $A$ is prevalent. Initially, nodes can be either
susceptible ($S$) or in the $A$ state in which $A$ has been adopted. We assume an initial fraction
$\rho_A$ of nodes in state $A$ and an initial fraction $\rho_S$ of nodes are in state $S$.
To model the early adoption of a new technology $B$, we also select a small initial fraction $\rho_B$ of
nodes in state $B$ as seeds to initiate the spreading of $B$.
We consider three different models for the mechanism of spreading of $B$: independent [Fig.~\ref{fig:model}(a)],
exclusive [Fig.~\ref{fig:model}(b)], and compatible [Fig.~\ref{fig:model}(c)] models.
i) In the independent model, the preexisting technology $A$ and the new technology $B$
are independent and do not interact with each other. If a fraction of neighbors of node
$i$ in $B$ state is larger than a threshold $\theta_B$ ($k_B^{i}/k^i \ge \theta_B$)
where $k_j^i$ is the number of neighbors of $i$ in $j$ state and $k^i$ is
the total number of neighbors $i$ (degree of node $i$), node $i$ changes to state $B$ state regardless
of the initial state of node $i$, either susceptible $S$ or $A$.
Therefore, the independent model is essentially the same as the original threshold
model for the spreading of a single technology \cite{granovetter,watts}. ii) In the exclusive
model, $A$ and $B$ compete and additional requirements exist for a user of $A$ to change its
state adopting $B$. To be specific, if $k_B^{i}/k^i \ge \theta_B$,
node $i$ in state $S$ adopts $B$, as the independent model. However, if node $i$ is in
state $A$, it changes to state $B$ only if the fraction of $B$ neighbors
of $i$ is larger than $\theta_B$ and at the same time the fraction of $A$ neighbors
is smaller than $\theta_A$ ($k_B^{i}/k^i \ge \theta_B$ and $k_A^{i}/k^i < \theta_A$).
iii) In the compatible model, we introduce dual users in an intermediate state $AB$ in the
transition from $A$ to $B$ modeling agents that use both technologies $A$ and $B$.
Nodes in the susceptible state $S$ change to state
$B$ as in the independent or exclusive model. However, a node $i$  in state $A$ becomes a dual user $AB$ by
adopting $B$ when $(k_B^{i}+k_{AB}^{i})/k^i \ge \theta_B$. A dual user $i$ ($AB$ state)
changes into $B$ state by discarding $A$, when the fraction of
neighbors in $A$ is less than the threshold for $A$ $(k_A^{i}/k^i < \theta_A)$.
In all three models, dynamics proceeds from the initial state until a final frozen configuration
is reached which depends on parameters such as threshold parameters $\theta_A$, $\theta_B$,
initial fraction $\rho_A$, $\rho_B$, and network structure as well.

\section*{Theory}
The general question addressed is about the conditions for the spreading of $B$, but in
particular we ask about the possible coexistence of $A$ and $B$ and the role of dual users
in the spreading of $B$. To compute the final fraction of nodes in each state we provide
a general theory of
complex contagion in which each node can be in $n$ states. The
initial fraction of nodes with state $x$ is $\rho_x$ where $x \in \{1,2,\cdots,n\}$.
The final average fraction of nodes in state $x$, $R_x$, can be expressed as
\begin{align}
\label{mx}
R_x=\sum_{y=1}^n \left[\rho_y f(y,x) - \rho_x f(x,y)\right],
\end{align}
where $f(x,y)$ is the transition probability from state $x$ to $y$.
On locally tree-like networks in the limit $N \rightarrow \infty$, the
transition probability from $x$ to $y$ can be computed as~\cite{watts,gleeson-seed,newman2},
\begin{align}
\label{mxy}
f(x,y)=\sum_{\vec{k}} P(k) T_0(\vec{k}) \prod_{w=1}^n q_w(x)^{k_w} \eta(x,y)
\end{align}
where $\eta(x,y)$ is a threshold function taking values $0$ or $1$ for the transition
from state $x$ to $y$ defined by the adoption rule for a pair of states $(x,y)$,
$k_x$ represents the degree of a node in a given state $x$, $k$ is the degree of a node
$k=\sum_{x \in \{1,2,\cdots,n \}} k_x$,
$P(k)$ is the degree distribution for nodes,
the vector of degree for each state is $\vec{k}=\{ k_1,k_2,\cdots,k_n \}$,
and multinomial distribution of them is
$T_0(\vec{k})=\frac{\prod_w k_w!}{k!}$.
Here, $q_x(y)$ is the probability that a node, connected to a node in state $y$, is in state $x$.

\begin{figure}
\begin{center}
\includegraphics[width=0.8\linewidth]{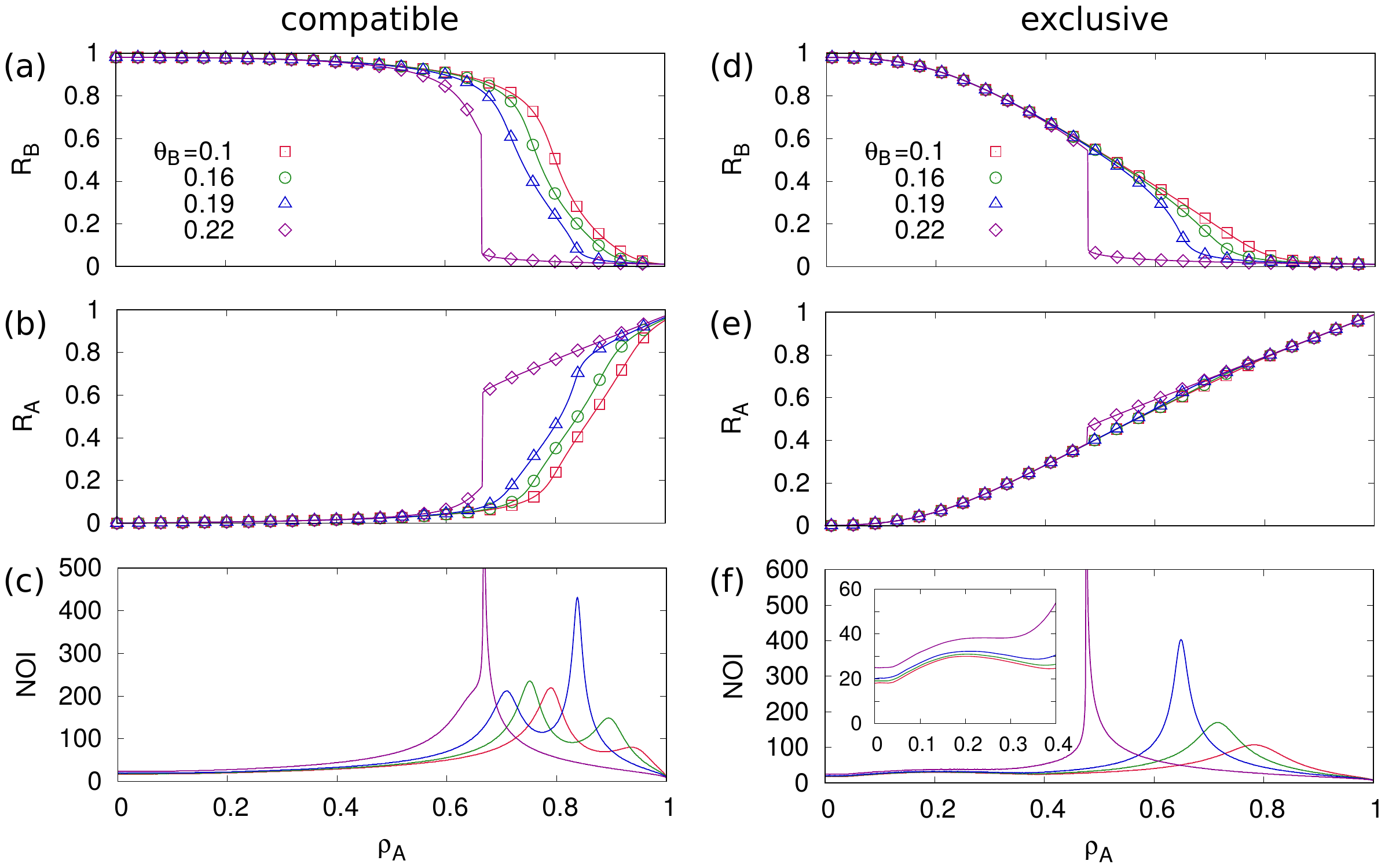}
\caption{
Theory (line) and simulation results (symbol) for the final fraction of $A$, $B$ and
NOI for (a-c) compatible and (d-f) exclusive
with respect to $\rho_A$ on ER networks with $N=10^5$, $\langle k \rangle =4$, $\theta_A=0.2$, and
$\rho_B=0.01$ are shown with $\theta_B=0.1$ $(\square)$, $0.16$ $(\circ)$,
$0.19$ $(\triangle)$, and $0.22$ $(\diamond)$. The local peak of NOI
accurately indicates the transition of the spread of $R_B$ and
the extinction of $R_A$.
}
\label{fig:fig2}
\end{center}
\end{figure}

In order to obtain $f(x,y)$, we first calculate $q_x(y)$ by introducing the transition
probability $g(x,y,z)$ from state $x$ to $y$ for a node connected to a node in state $z$:
\begin{align}
\label{meqq}
q_x(y)=\sum_{w=1}^n \left[\rho_w g(w,x,y) - \rho_x g(x,w,y)\right].
\end{align}
Next, we define the excess degree of a node for a given state $x$ as $\kappa_x$.
Considering $g(x,y,z)$ where a node is connected to a node in state $z$, we can
define the vector of excess degree for each state as
$\vec{\kappa}_z=\{ k_1, k_2, \cdots, k_z-1,\cdots,k_n \}$.
Assuming locally-tree like structures, we can compute $g(x,y,z)$
using the self-consistency equations~\cite{watts,gleeson-seed}:
\begin{align}
\label{mzxy}
	g(x,y,z)=&\sum_{\vec{\kappa_z}} \frac{k P(k)}{\langle k \rangle} T_1(\vec{\kappa}_z)
	\prod_{w=1}^n q_w(x)^{\kappa_w} \eta(x,y,z),
\end{align}
where $\eta(x,y,z)$ is a threshold function for the transition
from state $x$ to $y$ for a node connected to a node in state $z$ and
$T_1(\vec{\kappa}_z)=\frac{\prod_w \kappa_w!}{(k-1)!}$.
The differences among
possible threshold contagion models are encoded in the threshold functions $\eta(x,y)$
and $\eta(x,y,z)$ that account for different transition rules. Details of this theoretical
framework for our independent, exclusive, and compatible models are given in Methods.
Our approximation is exact in a tree structure and produces very good agreement
with numerical simulations for sparse tree-like graphs.

\section*{Results}

In a threshold spreading model of a single technology ($n=2$, $x=S$ or $x=A$),
there exists a single well known transition \cite{granovetter,watts} from a non-adopting
phase to a spreading phase in which, by a global cascade mechanism, a large fraction of
agents adopt the technology $x=A$ from an initial seed. In our set-up with $A$ initially
spread, and an initial seed of $B$, we search for transitions for the spreading of $B$,
as well as for the extinction of $A$. When these transitions do not coincide there will
be a phase of coexistence of $A$ and $B$. Different phases and transitions can be identified
calculating from Eqs. (1-4) the final fraction of nodes $R_A$ and $R_B$ in states $A$ and
$B$ respectively. We define four possible phases.
i) Phase \textbf{S}: extinction of $A$ and no spreading of $B$,
ii) Phase \textbf{A}: survival of $A$ and no spreading of $B$,
iii) Phase \textbf{B}: extinction of $A$ and spreading of $B$, and
iv) Coexistence phase \textbf{A+B}: survival of $A$ and spread of $B$.

\begin{figure}[t]
\begin{center}
\includegraphics[width=0.7\linewidth]{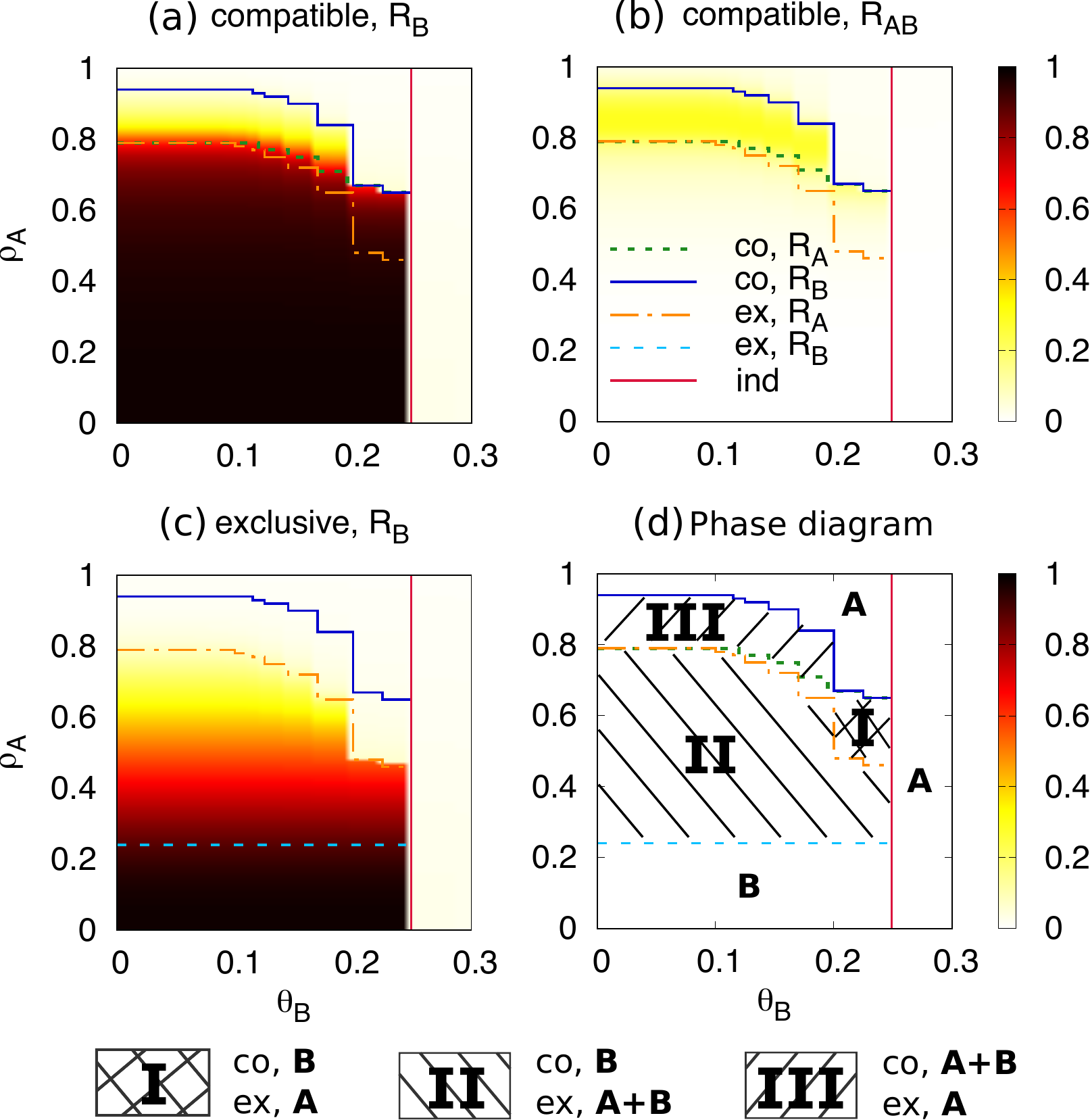}
\caption{
(a) $R_B$, (b) $R_{AB}$ for the compatible model, (c) $R_B$ for the exclusive
model, and (d) phase diagram in the $\rho_A$ - $\theta_B$ plane, with
$\langle k \rangle =4$, $\theta_A=0.2$, $\rho_A=0.6$, and $\rho_B=0.01$.
Transition lines are indicated as: i) Compatible model: Extinction of $A$
(co, $R_A$), spreading of $B$ (co, $R_B$), ii) Exclusive model: Extinction
of $A$ (ex, $R_A$), spreading of $B$ (ex, $R_B$), and
iii) Transition for the independent model (ind).
}
\label{fig:fig3}
\end{center}
\end{figure}

An example of how these phases can be identified for the compatible and exclusive
model is given in Fig.~\ref{fig:fig2} showing $R_A$ and $R_B$ for particular parameter values
in an Erd\"os-R\'enyi (ER) network with $\langle k \rangle =4$. The values chosen for $\theta_A$ and $\theta_B$
are in the range ($\theta<1/4$) in which there is a spreading phase in the
independent model.
Extremely good agreement is found between theory and simulation. The transition between
different phases can be accurately identified by a local maximum in the number of
iteration (NOI) of the recursion equations (Eq.~\ref{mzxy}), reflecting critical
slowing down at the transition point.
As the initial fraction of nodes in state $A$, $\rho_A$, decreases
we can observe two transitions in $R_A$ and $R_B$: a transition to extinction
of $A$ and a transition to spreading of $B$.
For the compatible model, the transitions
observed for $R_A$ and $R_B$ coincide when $\theta_B=0.22$ (single peak of NOI),
but they appear at different values of  $\rho_A$ when $\theta_B<0.2$ allowing for
a coexistence phase \textbf{A+B}.
Note also that these transitions are discontinuous for
$\theta>0.2$ ($\theta_B=0.22$), but become continuous for $\theta_B<0.2$.
For the exclusive model,
a coexistence phase (\textbf{A+B}) appears consistently for all $\theta_B$.
The spreading transition for $R_B$ is also discontinuous here for $\theta_B>0.2$
and continuous for $\theta_B<0.2$.

\begin{figure*}[t]
\begin{center}
\includegraphics[width=1\linewidth]{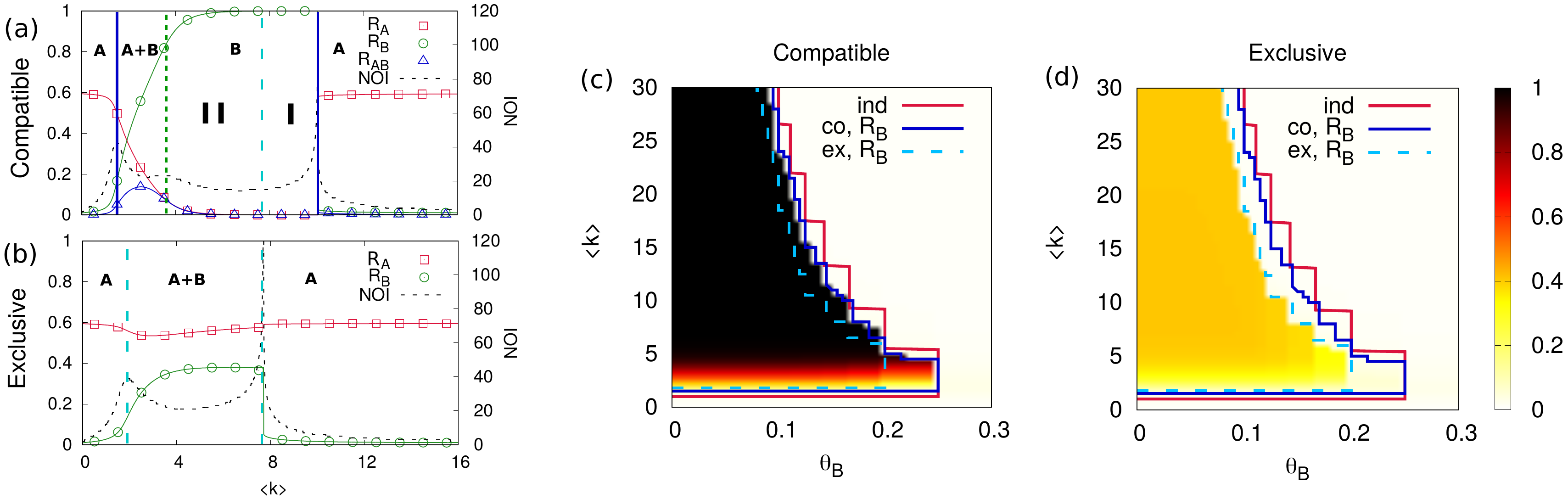}
\caption{
Theory (line) and numerical results (symbol) for the final fraction of
$A$, $B$, $AB$ nodes, and NOI for (a) compatible and (b) exclusive models on
ER networks ($N=10^5$) as a function of $\langle k \rangle$ with $\theta_A=0.2$, $\theta_B=0.16$,
$\rho_A=0.6$, and $\rho_B=0.01$. The phases (\textbf{A}, \textbf{B}, and \textbf{A+B}) and transitions
between them are identified by the peak of NOI. $R_B$ for (c) compatible and (d)
exclusive models in the $\langle k \rangle$ - $\theta_B$ plane, with $\theta_A=0.2$, $\rho_A=0.6$,
and $\rho_B=0.01$. Transition lines are indicated as in Fig.~\ref{fig:fig3}.
}
\label{fig:fig4}
\end{center}
\end{figure*}

Proceeding along these lines we can construct from our analytical solution phase diagrams 
as a function of different parameters:
Fig.~\ref{fig:fig3} shows a plot density of $R_B$ for the exclusive and compatible models in the plane
$\theta_B$ - $\rho_A$ with different transition lines (see Methods for their calculation).
Varying $\theta_B$, the transition lines in the phase diagram is discontinuous rather
than a continuous curve because the threshold of our model is based on the ratio to the degree.
Specifically, we can observe sharp transitions at the point where $\theta_B=n/k$ with $n$ being an integer.
Note that such discontinuity is also found in the prototypical threshold model \cite{watts}.
Phase \textbf{A} always exists for $\theta_B>1/4$ larger than the critical value of the independent
model ($\theta_c=1/4$) since $\theta_A$ is less than $1/4$. For $\theta_B<1/4$ phase \textbf{A} occurs also in
both models when $\rho_A$ is above the critical value for spreading of $B$ in the compatible
model. On the other hand, also for $\theta_B<1/4$, phase \textbf{B} occurs in both models when
$\rho_A$ is below the transition line for spreading of $B$ in the exclusive model.
In between these two regimes we find three domains of parameters in which results are
different for the two models. These domains give evidence of the important role of dual users
AB in favoring the spread of $B$. In region \textbf{I} we find phase \textbf{A} in the exclusive
model and phase \textbf{B} in the compatible model: This is an extreme effect of dual users leading
to the extinction of $A$ and spreading of $B$. In region \textbf{II} we find phase \textbf{A+B}
in the exclusive model and phase \textbf{B} in the compatible model: Dual users destroy coexistence
in favor of full spreading of $B$. In region \textbf{III} we find phase \textbf{A} in the exclusive
model and phase \textbf{A+B} in the compatible model: Dual users spread $B$ allowing for coexistence.
In this coexistence regime, there is a large proportion $R_{AB}$ of dual users in the final
state, while in regions \textbf{I} and \textbf{II} dual users play an important transient role
in the spreading of $B$, but essentially disappear in the final state. In summary, dual users
behave as an intermediate step in the transition from $A$ to $B$, mitigating the first-mover
advantages of $A$.

We next look at network effects that confirm the important role of dual users
in the spreading of $B$ (Fig.~4). We consider a point in region \textbf{II} of Fig.~\ref{fig:fig3}
and examine the effect of varying the average degree $\langle k \rangle$ of the ER network.
The compatible model shows three transitions when increasing $\langle k \rangle$. Since networks do
not form a globally connected component below the percolation threshold ($\langle k \rangle_{pc}=1$),
spreading of $B$ does not occur when $\langle k \rangle<1$. A first continuous spreading transition
for $B$ leads from phase \textbf{A} to the coexistence phase \textbf{A+B}.
A second continuous extinction transition for $A$ occurs around $\langle k \rangle \approx 4$ leading
from phase \textbf{A+B} to phase \textbf{B}. A final transition from \textbf{B} to \textbf{A}
occurs for the high connectivity regime ($\langle k \rangle \gtrsim 10$) where initial fraction of $B$ does
not exceed the threshold $\theta_B$, so that the spread of $B$ disappears abruptly.
For the exclusive model, phase \textbf{B} between \textbf{A+B} and \textbf{A} no longer
exists and $R_A>0.5$ for all values of $\langle k \rangle $, indicating a strong preemptive advantage of $A$.
Comparing the two models we find again regions \textbf{I} and \textbf{II} with the same
characteristics than in Fig.~\ref{fig:fig3} where dual users play a dominant role in the
spreading of $B$. Again this transient dynamics affects with only a significant number of
dual users in the final state ($R_{AB}$) in the coexistence phase \textbf{A+B} of the compatible model.

Dependence of our results on $\theta_B$ is seen in Fig.~\ref{fig:fig4}(c,d) where results
for $R_B$ are shown in the $\langle k \rangle$ - $\theta_B$ plane. The range of parameters in which $B$
spreads in the exclusive model is smaller than in the independent model due to the hindering
effect of a preexisting technology $A$, but this effect is smaller for the compatible model.
More importantly, when $B$ spreads, the final fraction $R_B$ is much larger in the compatible
model than in the exclusive model. Note that phase \textbf{B} in Fig.~\ref{fig:fig4}(a) is within
the $R_B$ transition line of the compatible model [Fig.~\ref{fig:fig4}(c)], while phase \textbf{A+B}
in Fig.~\ref{fig:fig4}(b) is within the $R_B$ transition line of the exclusive model
[Fig.~\ref{fig:fig4}(d)].
We also find that our main conclusions are consistently valid for a broad range of
parameters.

\section*{Discussion}

In summary, we have discussed models of complex contagion for the spreading of a
new innovation $B$ competing with an established one $A$. We obtain perfect agreement
between theory and simulation. Competition, as compared with the independent model, is
shown to have a hindering effect on the spreading of the new technology. However, competition
can create a gap between the extinction and spreading transitions that allows for
a coexistence phase. It can also change the nature of the spreading transition from
first to second order. Moreover, the existence of dual users is found to be a key
mechanism to promote the spreading of $B$ in a system dominated by a $A$,
allowing to overcome the hindering effect of tightly clustered of agents using $A$.
Dual users create new phases in which they either lead to extinction of $A$ in
favor of spreading of $B$, or destroy coexistence in favor of spreading of $B$
or create coexistence by $B$ spreading.
The role of dual users provides a plausible
explanation to reconcile two seemingly contradictory factors: the first-mover
advantage meaning that the prevalence of $A$ potentially blocks
the spread of $B$, and the success of late-mover frequently observed in reality.
Practically, our study suggests a strategic way of spreading a new technology
in a competitive environment by offering it as a good compatible alternative
instead of promoting changing to the new technology.
Further studies may be needed to examine the role of dual-users
in simple contagion models, opinion dynamics, and coevolutionary dynamics,
to name a few.

\section*{Methods}

\subsection*{Theory of compatible model}
We implement the developed general theory of complex contagion into the compatible model.
The final fraction of state $A$, $B$, $AB$, and $S$ are defined as $R_A$, $R_B$, $R_{AB}$,
and $R_S$, respectively. We further define the initial fraction of $A$, $B$, $AB$, and $S$
as $\rho_A$, $\rho_B$, $\rho_{AB}$ and $\rho_S$. These variables are normalized as
$R_A+R_B+R_{AB}+R_S=1$ and $\rho_A+\rho_B+\rho_{AB}+\rho_S=1$.
Note that $\rho_{AB}=0$ since the coexisting state cannot exist initially.
Then the main purpose of the theory is to calculate the final fraction of nodes
in each state for a given initial condition. Given the initial fraction of each state,
the final fraction of each state for the compatible model is given by
\begin{align}
\label{final1}
R_A &= \rho_A [1-f(A,AB)-f(A,B)],\\
\label{final2}
R_B &= \rho_B + \rho_S f(S,B) + \rho_A f(A,B),\\
\label{final3}
R_{AB} &= \rho_A f(A,AB),
\end{align}
where $f(x,y)$ represents the transition probability from state $x$ to $y$.
Then, the transition probability $f(x,y)$ on a locally-tree like network can be
calculated as
\begin{align}
\label{fxy}
f(x,y)&=\sum_{\vec{k}} P(k) T_0(\vec{k}) q_A(x)^{k_A} q_B(x)^{k_B}
q_{AB}(x)^{k_{AB}} q_S(x)^{k_S} \eta(x,y).
\end{align}
Here, $k_x$ represents degree of a node for a given state $x$, $k$ is the degree of a node
$k=\sum_{x \in \{A,B,{AB},S\}} k_x$,
$P(k)$ is the degree distribution of underlying networks,
the vector of degree for each state is $\vec{k}=(k_S,k_A,k_B,k_{AB})$,
and multinomial distribution of them is
$T_0(\vec{k})=\frac{k_A! k_B! k_{AB}! k_S!}{k!}$.
$q_x(y)$ is the probability that a node, connected to a node in state $y$, is in state $x$.
And, threshold $\eta(x,y)$ is given by the rule of adoption for a pair of state $(x,y)$.
Specifically
$\eta(S,B)=H(\frac{k_B+k_{AB}}{k}-\theta_B)$,
$\eta(A,B)=H(\frac{k_B}{k}-\theta_B) H(\theta_A - \frac{k_A}{k})$, and
$\eta(A,AB)=H(\frac{k_B}{k}-\theta_B) H(\frac{k_A}{k}-\theta_A)$
where $H(x)$ is the Heaviside step function, $H(x)=0$ if $x<0$ and $H(x)=1$ if $x \ge0$.

A set of variables $q_x(y)$ can be obtained by introducing the transition
probability $g(x,y,z)$ from state $x$ to $y$ for a node connected to a node in state $z$:
\begin{align}
\label{eqq1}
q_A(z) &= \rho_A [1 - g(A,AB,z) - g(A,B,z)], \\
\label{eqq2}
q_B(z) &= \rho_B + \rho_S g(S,B,z) + \rho_A g(A,B,z), \\
\label{eqq3}
q_{AB}(z) &= \rho_A g(A,AB,z),
\end{align}
where $z \in \{ S,A \}$. Furthermore, $g(x,y,z)$ can be calculated
by the self-consistency equations with the locally-tree like structures
\begin{align}
\label{fzxy}
	g(x,y,z)&=\sum_{\vec{\kappa_z}} \frac{k P(k)}{\langle k \rangle}  T_1(\vec{\kappa_z})
	q_A(x)^{\kappa_A} q_B(x)^{\kappa_B} q_{AB}(x)^{\kappa_{AB}}
	q_S(x)^{\kappa_S} \eta(x,y,z).
\end{align}
where $\kappa_i$ is the excess degree of state $i$, and
$\eta(x,y,z)$ is a threshold function for the transition
from state $x$ to $y$ for a node connected to a node in state $z$.
$\eta(x,y,z)$ for our compatible model is given by
$\eta(S,B,S)=\eta(S,B,A)=H(\frac{k_B+k_{AB}}{k}-\theta_B)$,
$\eta(A,B,S)=H(\frac{k_B}{k}-\theta_B) H(\theta_A - \frac{k_A}{k})$,
$\eta(A,B,A)=H(\frac{k_B}{k}-\theta_B) H(\theta_A - \frac{k_A+1}{k})$,
$\eta(A,AB,S)=H(\frac{k_B}{k}-\theta_B) H(\frac{k_A}{k} - \theta_A)$. and
$\eta(A,AB,A)=H(\frac{k_B}{k}-\theta_B) H(\frac{k_A+1}{k} - \theta_A)$.
Note that $\eta(A,B,A)$ is different from $\eta(A,B,S)$ in
$H(\theta_A - \frac{k_A+1}{k})$ since in $\eta(A,B,A)$ an additional neighbor with $A$
state should be considered.

\subsection*{Theory of exclusive model}

For the exclusive model, since coexisting $AB$ state is not
allowed, the final fraction of each state is given by
\begin{align}
R_A &= \rho_A [1-f(A,B)],\\
R_B &= \rho_B + \rho_S f(S,B) + \rho_A f(A,B).
\end{align}
Similarly, the variables $q_x(y)$ are
\begin{align}
q_A(z) &= \rho_A [1 -  g(A,B,z)], \\
q_B(z) &= \rho_B + \rho_S g(S,B,z) + \rho_A g(A,B,z),
\end{align}
where $z \in \{ S,A \}$.
The transition probability $f(x,y)$ and $g(x,y,z)$ can be computed
by using the same equations for compatible model (Eqs.~\ref{fxy} and \ref{fzxy})
but with different $\eta(x,y)$ and $\eta(x,y,z)$.
To be specific, for the exclusive model,
$\eta(S,B)=H(\frac{k_B}{k}-\theta_B)$,
$\eta(A,B)=H(\frac{k_B}{k}-\theta_B) H(\theta_A - \frac{k_A}{k})$,
$\eta(S,B,S)=\eta(S,B,A)=H(\frac{k_B}{k}-\theta_B)$,
$\eta(A,B,S)=H(\frac{k_B}{k}-\theta_B) H(\theta_A - \frac{k_A}{k})$, and
$\eta(A,B,A)=H(\frac{k_B}{k}-\theta_B) H(\theta_A - \frac{k_A+1}{k})$.

\subsection*{Theory of independent model}

For the independent model, nodes in $A$ state do not interact with nodes in
$B$ state and thus the final fraction of nodes is given by
\begin{align}
R_A &= \rho_A [1-f(A,B)],\\
R_B &= \rho_B + \rho_S f(S,B) + \rho_A f(A,B).
\end{align}
And $q_x(y)$ are
\begin{align}
q_A(z) &= \rho_A [1 -  g(A,B,z)], \\
q_B(z) &= \rho_B + \rho_S g(S,B,z) + \rho_A g(A,B,z),
\end{align}
where $z \in \{ S,A \}$.
The thresholds for the independent model are simply given by
$\eta(S,B)=\eta(A,B)=H(\frac{k_B}{k}-\theta_B)$ and
$\eta(S,B,S)=\eta(S,B,A)=\eta(A,B,S)=\eta(A,B,A)=H(\frac{k_B}{k}-\theta_B)$.

\section*{Acknowledgements}
We acknowledge financial support from Agencia Estatal de Investigacion (AEI, Spain) and 
Fondo Europeo de Desarrollo Regional under Project ESoTECoS Grant No.FIS2015-63628-C2-2-R (AEI/FEDER,UE) 
and the Spanish State Research Agency, through the Maria de Maeztu Program 
for Units of Excellence in R\&D (MDM-2017-0711).
This work was supported by the research grant of the Chungbuk National
University in 2018.

\section*{Author Contributions}
All authors conceived the study,
B. M. performed the research,
all authors analyzed the data, and
wrote and reviewed the paper.

\section*{Additional Information}
Competing Interests: The authors declare no competing interests.



\begin{thebibliography}{99}



\bibitem{geroski} Geroski, P. A.
	Models of technology diffusion.
	Research policy {\bf 29}, 603-625 (2000).

\bibitem{karsai} Karsai, M., I\~niguez, G., Kaski, K. \& Kert\'esz, J.
	Complex contagion process in spreading of online innovation.
	J. R. Soc. Interface {\bf 11}, 20140694 (2014).

\bibitem{schelling} Schelling, T. C.
	Hockey helmets, concealed weapons, and daylight saving: A study of binary choices with externalities.
	J. Conflic Resolut. {\bf 17}, 381 (1973).

\bibitem{granovetter} Granovetter, M.
	Threshold models of collective behavior.
	Am. J. Sociol, {\bf 83}, 1420 (1978).

\bibitem{watts} Watts, D. J.
	A simple model of global cascades on random networks.
	Proc. Natl. Acad. Sci. {\bf 99}, 5766-5771 (2002).

\bibitem{centola} Centola, D.
	The spread of behavior in an online social network experiment.
	Science {\bf 329}, 1194 (2010).

\bibitem{gai} Gai, P. \&  Kapadia, S.
	Contagion in financial networks.
	Proc. R. Soc. A {\bf 466}, 2401 (2010).

\bibitem{dodds} Dodds, P. S. \& Watts, D. J.
	Universal behavior in a generalized model of contagion.
	Phys. Rev. Lett. {\bf 92}, 218701 (2004).

\bibitem{christakis} Christakis, N. A. \& Fowler, J. H.
	Social contagion theory: examining dynamics social networks and human behavior.
	Statist. Med. {\bf 32}, 556 (2013).

\bibitem{singh} Singh, P., Sreenivasan, S., Szymanski, B. K. \& Korniss, G.
	Threshold-limited spreading in social networks with multiple initiators.
	Sci. Rep. {\bf 3}, 2330 (2013).

\bibitem{ruan} Ruan, Z., I\~niguez, G., Karsai, M. \& Kert\'esz, J.
	Kinetics of social contagion.
	Phys. Rev. Lett. {\bf 115}, 218702 (2015).

\bibitem{macy} Centola, D., Egu\'iluz, V. M. \& Macy, M. W.
	Cascade dynamics of complex propagation.
	Physica A {\bf 374}, 449 (2007).

\bibitem{weng-vir} Weng, L., Menczer, F. \& Ahn, Y.-Y.
	Virality prediction and community structure in social networks.
	Sci. Rep. {\bf 3}, 2522 (2013).

\bibitem{monsted} Monsted, B., Sapiezynski, P., Ferrara, E. \& Lehman, S.
	Evidence of complex contagion of information in social media: An experiment using twitter bots.
	PLoS ONE {\bf 12(9)}, e0184148 (2017).

\bibitem{min-maxi} Min, B. \& San Miguel, M.
	Competing contagion processes: Complex contagion triggered by simple contagion.
	Sci. Rep. {\bf 8(1)}, 10422 (2018).


\bibitem{bailey} Bailey, N.
	{\it The mathematical theory of infectious diseases and its applications}, 2nd ed.
	(Griffin, London, 1975).

\bibitem{gleeson-seed} Gleeson, J. P. \& Cahalane, D. J.
	Seed size strongly affects cascades on random networks.
	Phys. Rev. E {\bf 75}, 056103 (2007).

\bibitem{melnik-tree} Melnik, S., Hackett, A., Porter, M. A., Mucha, P. J. \& Gleeson, J. P.
	The unreasonable effectiveness of tree-based theory for networks with clustering.
	Phys. Rev. E {\bf 83}, 036112 (2011).

\bibitem{galstyan} Galstyan, A. \& Cohen, P.
	Cascading dynamics in modular networks.
	Phys. Rev. E {\bf 75}, 036109 (2007).

\bibitem{gleeson-modular} Gleeson, J. P.
	Cascades on correlated and modular random networks.
	Phys. Rev. E {\bf 77}, 046117 (2008).

\bibitem{nematzadeh} Nematzadeh, A., Ferrara, E., Flammini, A. \& Ahn, Y.-Y.
	Optimal network modularity for information diffusion,
	Phys. Rev. Lett. {\bf 113}, 088701 (2014).

\bibitem{brummitt} Brummitt, C. D., Lee, K.-M. \& Goh, K.-I.
	Multiplexity-facilitated cascades in networks.
	Phys. Rev. E {\bf 85}, 045102(R) (2012).

\bibitem{lee} Lee, K.-M., Brummitt, C. D. \& Goh, K.-I.
	Threshold cascades with response heterogeneity in multiplex networks.
	Phys. Rev. E {\bf 90}, 062816 (2014).

\bibitem{melnik} Melnik, S., Ward, J. A., Gleeson, J. P. \& Porter, M. A.
	Multi-stage complex contagion.
	Chaos {\bf 23}, 013124 (2013).

\bibitem{aga} Czaplicka, A., Toral, R. \& San Miguel, M.
	Competition of simple and complex adoption on interdependent networks.
	Phys. Rev. E {\bf 94}, 062301 (2016).

\bibitem{lugo} Lugo, H. \& San Miguel, M.
	Learning and coordinating in a multilayer network.
	Sci. Rep. {\bf 5}, 7776 (2015).

\bibitem{messenger} https://ondeviceresearch.com/blog/messenger-wars-how-facebook-climbed-number-one

\bibitem{sns} http://www.pewinternet.org/2015/01/09/social-media-update-2014/

\bibitem{cyworld} https://www.techinasia.com/cyworld-vs-facebook/


\bibitem{kocsis} Kocsis, G. \& Kun, F.
	Competition of information channels in the spreading of innovations.
	Phys. Rev. E {\bf 84}, 026111 (2011).

\bibitem{newman} Newman, M. E. J.
	Threshold effects for two pathogens spreading on a network.
	Phys. Rev. Lett {\bf 95}, 108701 (2005).

\bibitem{ahn} Ahn, Y.-Y., Jeong, H., Masuda, N. \& Noh, J. D.
	Epidemic dynamics of two species of interacting particles on scale-free networks.
	Phys. Rev. E {\bf 74}, 066113 (2006).

\bibitem{karrer} Karrer, B. \& Newman, M. E. J.
	Competing epidemics on complex networks.
	Phys. Rev. E {\bf 84}, 036106 (2011).


\bibitem{castello} Castell\'o, X., Egu\'iluz, V. M. \& San Miguel, M.
	Ordering dynamics with two non-excluding opinions: bilingualism in language competition.
	New J. Phys. {\bf 8}, 308 (2006).

\bibitem{carro} Carro, A., Toral, R. \& San Miguel, M.
	Coupled dynamics of node and link states in complex networks: A model for language competition.
	New J. Phys. {\bf 18}, 113056 (2016).

\bibitem{amato} Amato, R., Kouvaris, N., San Miguel, M. \& D\'iaz-Guilera, A.
	Opinion competition dynamics on multiplex networks.
	New J. Phys. {\bf 19}, 123019 (2017).

\bibitem{chen} Chen, L., Ghanbarnejad, F., Cai, W. \& Grassberger, P.
	Outbreaks of coinfections: the critical role of cooperativity.
	EPL {\bf 104}, 50001 (2013).

\bibitem{cai} Cai, W., Chen, L., Ghanbarnejad, F. \& Grassberger, P.
	Avalanche outbreaks emerging in cooperative contagion.
	Nat. Phys. {\bf 11}, 936-940 (2015).

\bibitem{choi} Choi, W., Lee, D. \& Kahng, B.
	Critical behavior of a two-step contagion model with multiple seeds.
	Phys. Rev. E {\bf 95}, 062115 (2017).


\bibitem{weng} Weng, L., Flammini, A., Vespignani, A. \& Menczer, F.
	Competition among memes in a world with limited attention.
	Sci. Rep. {\bf 2}, 355 (2012).

\bibitem{sahneh} Sahneh, F. D. \& Scoglio, C.
	Competitive epidemic spreading over arbitrary multilayer networks.
	Phys. Rev. E {\bf 89}, 062817 (2014).

\bibitem{granell} Granell, C., G\'omez, S. \& Arenas, A.
	Dynamical interplay between awareness and epidemic spreading in multiplex networks.
	Phys. Rev. Lett. {\bf 111}, 128701 (2013).

\bibitem{min} Min, B. \& Goh, K.-I.
	Multiple resource demands and viability in multiplex networks.
	Phys. Rev. E {\bf 89}, 040802(R) (2014).

\bibitem{sanz} Sanz, J., Xia, C.-Y., Meloni, S. \& Moreno, Y.
	Dynamics of interacting diseases.
	Phys. Rev. X {\bf 4}, 041005 (2014).

\bibitem{newman2} Newman, M. E. J., Strogatz, S. H. \& Watts, D. J.
	Random graphs with arbitrary degree distributions and their applications.
	Phys. Rev. E {\bf 64}, 026118 (2001).

\end{thebibliography}
\end{document}